\documentclass[aps,prx,twocolumn,superscriptaddress]{revtex4}

\usepackage[dvips]{graphicx}
\usepackage{color}
\usepackage{bm}
\usepackage{amssymb}
\renewcommand{\vec}[1]{\boldsymbol{#1}}
\usepackage[utf8]{inputenc}
\usepackage{times}

\begin{document}

\title{Revisiting spin ice physics in the ferromagnetic Ising pyrochlore Pr$_2$Sn$_2$O$_7$}

\author{Brenden~R.~Ortiz} \altaffiliation{\href{mailto:ortizbr@ornl.gov}{ortizbr@ornl.gov}}
\affiliation{Materials Science and Technology Division, Oak Ridge National Laboratory, Oak Ridge, TN 37831, USA}

\author{Paul~M.~Sarte}
\affiliation{Materials Department, University of California Santa Barbara, Santa Barbara, CA 93106, USA}

\author{Ganesh~Pokharel}
\affiliation{Materials Department, University of California Santa Barbara, Santa Barbara, CA 93106, USA}

\author{Miles~J.~Knudston}
\affiliation{Materials Department, University of California Santa Barbara, Santa Barbara, CA 93106, USA}

\author{Steven~J.~Gomez~Alvarado}
\affiliation{Materials Department, University of California Santa Barbara, Santa Barbara, CA 93106, USA}

\author{Andrew~F.~May}
\affiliation{Materials Science and Technology Division, Oak Ridge National Laboratory, Oak Ridge, TN 37831, USA}

\author{Stuart~Calder}
\affiliation{Neutron Scattering Division, Oak Ridge National Laboratory, Oak Ridge, TN 37831, USA}

\author{Lucile~Mangin-Thro}
\affiliation{Institut Laue-Langevin, 71 avenue des Martyrs, 38000 Grenoble, France}

\author{Andrew~R.~Wildes}
\affiliation{Institut Laue-Langevin, 71 avenue des Martyrs, 38000 Grenoble, France}

\author{Haidong~Zhou}
\affiliation{Department of Physics and Astronomy, University of Tennessee, Knoxville, TN 37996, USA}

\author{Gabriele~Sala} 
\affiliation{Spallation Neutron Source, Second Target Station, Oak Ridge National Laboratory, Oak Ridge, TN 37831, USA}

\author{Chris~R.~Wiebe} 
\affiliation{Department of Chemistry, University of Manitoba, Winnipeg, MB R3T 2N2, Canada}
\affiliation{Department of Chemistry, University of Winnipeg, Winnipeg, MB R3B 2E9, Canada}
\affiliation{Department of Physics and Astronomy, McMaster University, Hamilton, ON L8S 4M1, Canada}

\author{Stephen~D.~Wilson}
\affiliation{Materials Department and California Nanosystems Institute, University of California Santa Barbara, Santa Barbara, CA 93106, USA}

\author{Joseph~A.~M.~Paddison} \altaffiliation{\href{mailto:paddisonja@ornl.gov}{paddisonja@ornl.gov}}
\affiliation{Materials Science and Technology Division, Oak Ridge National Laboratory, Oak Ridge, TN 37831, USA}

\author{Adam~A.~Aczel} \altaffiliation{\href{mailto:aczelaa@ornl.gov}{aczelaa@ornl.gov}}
\affiliation{Neutron Scattering Division, Oak Ridge National Laboratory, Oak Ridge, TN 37831, USA}

\date{\today}

\begin{abstract}
Pyrochlore materials are characterized by their hallmark network of corner-sharing rare-earth tetrahedra, which can produce a wide array of complex magnetic ground states. Ferromagnetic Ising pyrochlores often obey the ``two-in-two-out" spin ice rules, which can lead to a highly-degenerate spin structure. Large moment systems, such as Ho$_2$Ti$_2$O$_7$ and Dy$_2$Ti$_2$O$_7$, tend to host a classical spin ice state with low-temperature spin freezing and emergent magnetic monopoles. Systems with smaller effective moments, such as Pr$^{3+}$-based pyrochlores, have been proposed as excellent candidates for hosting a ``quantum spin ice" characterized by entanglement and a slew of exotic quasiparticle excitations. However, experimental evidence for a quantum spin ice state has remained elusive. Here, we show that the low-temperature magnetic properties of Pr$_2$Sn$_2$O$_7$ satisfy several important criteria for continued consideration as a quantum spin ice. We find that Pr$_2$Sn$_2$O$_7$ exhibits a partially spin-frozen ground state with a large volume fraction of dynamic magnetism. Our comprehensive bulk characterization and neutron scattering measurements enable us to map out the magnetic field-temperature phase diagram, producing results consistent with expectations for a ferromagnetic Ising pyrochlore. We identify key hallmarks of spin ice physics, and show that the application of small magnetic fields ($\mu_0 H_c \sim$~0.75~T) suppresses the spin ice state and induces a long-range ordered magnetic structure. Together, our work clarifies the current state of Pr$_2$Sn$_2$O$_7$ and encourages future studies aimed at exploring the potential for a quantum spin ice ground state in this system. 
\end{abstract}

\maketitle

\section{Introduction}

Quantum spin liquids \cite{10_balents, 16_norman, 17_savary, 17_zhou, 19_knolle, 19_takagi, 19_wen, 20_broholm, 21_chamorro} and spin ices \cite{01_bramwell_2, 04_melko, 12_castelnovo, 20_bramwell, 21_jaubert} are some of the exotic states of matter that arise from the interplay of magnetism on a frustrated lattice. In both cases, frustration impedes the ability to satisfy competing exchange interactions, producing a largely degenerate ground state and suppressing conventional magnetic order. The rare earth pyrochlores exhibit a frustrated lattice of corner-sharing tetrahedra decorated with rare earth ions, and exhibit two different varieties of spin ice. Large moment systems such as Dy$_2$Ti$_2$O$_7$ and Ho$_2$Ti$_2$O$_7$ exhibit a classical spin ice (CSI) state, with well-isolated crystal field ground state doublets and nearly pure $|M_J = \pm J>$ wavefunctions, where $J$ is the total angular momentum quantum number. Here, the spins on each tetrahedron obey the ``two-in-two-out" ice rule and the thermal excitations behave as emergent magnetic charges \cite{08_castelnovo}. The quantum spin ice (QSI) state \cite{11_ross, 14_gingras, 21_jaubert}, which is a particular type of quantum spin liquid, expands on this phenomenology and supports a full emergent electrodynamics with different flavors of quasiparticles that resemble photons, electric charges, and magnetic charges. In principle, this novel state can be realized by adding quantum-tunnelling terms arising from significant quantum fluctuations to the CSI model \cite{12_benton, 13_savary}. In practice, this requires terms in the magnetic Hamiltonian that couple components of the spins transverse to the Ising axis. Although well-isolated crystal field ground state doublets are still important for satisfying this criterion, the wavefunctions must now contain a significant admixture of terms $|M_J \neq \pm J>$. Despite some success in identifying rare earth pyrochlores with the desired crystal field ground states, spin ice correlations, and significant quantum fluctuations, definitively establishing a QSI state in a real material has proven to be difficult \cite{14_gingras, 21_jaubert}. The search for quantum analogs of ice therefore remains a key goal of condensed matter physics.

Pyrochlores based on Pr$^{3+}$ ions ($J=4$) have been identified as intriguing QSI candidates. Of these, Pr$_2$Zr$_2$O$_7$ and Pr$_2$Hf$_2$O$_7$ have received the most attention due to the availability of large single crystals grown by the floating zone method \cite{14_hatnean, 14_koohpayeh, 17_hatnean, 18_anand}. The Pr$^{3+}$ crystal field ground states are non-Kramers doublets with Ising anisotropy \cite{13_kimura, 16_bonville, 16_anand, 16_sibille} and they are well-separated from the first excited levels with large energy gaps of 9.5 meV and 9.1 meV for the Zr \cite{13_kimura, 16_bonville} and Hf \cite{16_anand, 16_sibille} systems respectively. Accordingly, a $J{\rm{_{eff}}}=\frac{1}{2}$ pseudo-spin model can provide an appropriate description of the low-temperature magnetic properties. Evidence for spin ice correlations includes Pauling's residual entropy \cite{13_kimura}, pinch point scattering \cite{13_kimura, 16_petit, 18_sibille}, and a metamagnetic transition to a kagome ice state when the applied magnetic field $H \parallel [111]$ \cite{16_sibille, 22_tang}. The observation of quasi-elastic scattering below 1 K \cite{13_kimura, 16_sibille, 17_wen, 18_sibille} combined with evidence for partial spin freezing \cite{09_matsuhira, 13_kimura, 16_petit, 16_sibille, 16_anand} indicates that part of the spin system remains dynamic and therefore demonstrates the importance of quantum fluctuations for both materials \cite{13_kimura, 16_petit, 17_wen, 18_sibille}. This behavior is very different to that of CSIs such as Ho$_2$Ti$_2$O$_7$ and Dy$_2$Ti$_2$O$_7$, which show no measurable inelastic scattering in the same energy range \cite{08_clancy} and have much longer spin correlation times at low temperature \cite{03_synder, 04_ehlers}. Key signatures of the QSI state, including suppressed pinch point scattering and an inelastic spin excitation continuum, have been identified in Pr$_2$Hf$_2$O$_7$ \cite{18_sibille}. The unusual low-temperature heat conduction for Pr$_2$Zr$_2$O$_7$ has also been attributed to a QSI state, argued to arise from the different quasiparticles associated with it \cite{18_tokiwa}. Unfortunately, most Pr$_2$Zr$_2$O$_7$ samples are known to be affected by structural disorder likely arising from Pr off-centering or Pr$^{4+}$ defects \cite{14_koohpayeh}. While there is currently no consensus if this disorder is strong enough to induce a paramagnetic state with quadrupolar correlations \cite{16_petit, 17_martin, 18_benton} or if a quantum spin liquid state is realized instead \cite{17_wen}, recent success in synthesizing and characterizing pristine Pr$_2$Zr$_2$O$_7$ crystals suggests that the intrinsic ground state of this system is a QSI \cite{22_tang}.

Pr$_2$Sn$_2$O$_7$ has also been put forth as a QSI candidate \cite{08_zhou}. The Pr$^{3+}$ ions have a non-Kramers doublet ground state with an energy gap of 17.8 meV to the first excited state \cite{08_zhou, 13_princep}, which is significantly larger than for the Zr and Hf analogs. The powder-averaged magnetic diffuse scattering is reminiscent of expectations for CSI but not completely captured by this model, with improved agreement between theory and experiment obtained when considering an anisotropic exchange model with quantum fluctuations instead \cite{10_onoda, 11_onoda}. Commonalities with the other Pr-based pyrochlores include Ising moments \cite{13_princep}, a small, positive Curie-Weiss temperature \cite{08_zhou}, and quasi-elastic scattering below 1 K \cite{08_zhou} that are typically seen as strong evidence for spin ice correlations and significant quantum fluctuations. Key differences between Pr$_2$Sn$_2$O$_7$ and the other Pr pyrochlores have also been identified. A residual entropy greater than Pauling's value was reported in Pr$_2$Sn$_2$O$_7$, which was partially suppressed by a 1 T applied magnetic field \cite{08_zhou}. Previous AC susceptibility measurements of Pr$_2$Sn$_2$O$_7$ show evidence for the entire spin system freezing \cite{04_matsuhira}, which is in line with expectations for CSI, but apparently inconsistent with related work on Pr$_2$Sn$_2$O$_7$ \cite{08_zhou} and the other Pr pyrochlores, which generally support dynamical ground states with partial (or drastically suppressed) spin freezing signatures \cite{13_kimura, 16_petit, 16_sibille}. Another unique aspect of Pr$_2$Sn$_2$O$_7$ is the fine structure that was observed in its quasi-elastic neutron scattering spectrum. This was explained by a magnetic monopole confinement model appropriate for QSI arising from a finite tension of the Dirac strings that connect monopole pairs \cite{17_sarte}, although the origin for the linear potential giving rise to the string tension is still unknown. 

Despite the intriguing properties of Pr$_2$Sn$_2$O$_7$, the ongoing interest in QSI physics, and the notable discrepancies between some key properties of this material and the other Pr pyrochlores, subsequent investigations of Pr$_2$Sn$_2$O$_7$ have been conspicuously absent in the literature. We address this issue here by performing a series of comprehensive measurements aimed at characterizing the low-temperature magnetic properties of this QSI candidate in detail. We confirm key spin ice signatures including residual Pauling entropy and we illustrate the importance of quantum fluctuations in Pr$_2$Sn$_2$O$_7$ via bulk characterization measurements of our polycrystalline samples. Next, using our field-dependent measurements, we establish a $H$-$T$ phase diagram for Pr$_2$Sn$_2$O$_7$ that is consistent with a ferromagnetic Ising pyrochlore. We show that the spin ice state is extremely fragile, with long-range magnetic order achieved upon application of a modest magnetic field $\mu_0 H<$1~T. Finally, we establish a dynamical magnetic ground state for Pr$_2$Sn$_2$O$_7$ with a partial spin freezing signature, as has been identified in other Pr pyrochlores. Our combined results show that Pr$_2$Sn$_2$O$_7$ is a spin ice system that should receive continued consideration as a potential host of an exotic QSI state. 

\section{Experimental details}
\subsection{Synthesis}
Polycrystalline samples of Pr$_2$Sn$_2$O$_7$ were synthesized by the traditionally reported standard solid-state reaction using stoichiometric amounts of Pr$_6$O$_{11}$  (99.99\%, Alfa Aesar) and SnO$_2$  (99.99\%, Alfa Aesar). For consistency with prior literature methods, the powder reagents were mixed together and finely ground with a mortar and pestle. The resulting fine powder was placed in an alumina crucible and pre-reacted by heating in air at 1000$^{\circ}$C for 24~h. Samples were then reground and heated in air at 1400$^{\circ}$C for approximately 48~h with intermittent grindings until room temperature powder x-ray diffraction (XRD) measurements confirmed no discernible impurities. Phase purity was confirmed with room temperature XRD on a PANalytical X'Pert Pro MPD diffractometer (monochromated Cu K$_{\alpha 1}$ radiation) in standard Bragg-Brentano ($\theta$-2$\theta$) geometry. Rietveld refinements of the powder XRD patterns were performed using TOPAS ACADEMIC v6 \cite{TOPAS}. Structural models and visualization utilized the VESTA software package \cite{VESTA}. 

\subsection{Magnetization and Heat Capacity Measurements}
Temperature and field-dependent DC magnetization measurements were performed on a 7~T Quantum Design (QD) Magnetic Property Measurement System (MPMS3) SQUID magnetometer in vibrating-sample (VSM) mode. Polycrystalline Pr$_2$Sn$_2$O$_7$ was placed in a polypropylene capsule and subsequently mounted in a brass holder. The temperature and field-dependent AC magnetization measurements were performed on a QD 9~T Dynacool Physical Property Measurement System (PPMS) employing the AC susceptibility option for the dilution fridge (ACDR). Pr$_2$Sn$_2$O$_7$ powder was cold pressed with a Carver press, and a portion of the resulting pellet with approximate dimensions of 1 $\times$ 1 $\times$ 0.5~mm$^3$ was adhered to a sapphire sample mounting post with a thin layer of Apiezon N-grease. Unless specified, all AC susceptibility measurements were collected under field-cooled (FC) conditions. 

The heat capacity of Pr$_2$Sn$_2$O$_7$ and its corresponding nonmagnetic analog La$_2$Sn$_2$O$_7$ were measured with a QD 9~T Dynacool Physical Property Measurement System (PPMS) for both conventional $^{4}$He and dilution refrigerator temperature ranges. For the dilution fridge measurements, powder samples of Pr$_2$Sn$_2$O$_7$ and La$_2$Sn$_2$O$_7$ were blended with silver powder (Alfa Aesar, 99.999\%) to aid in thermal coupling to the sample stage and a more accurate assessment of the nuclear Schottky anomaly in the 60-500~mK range.

\subsection{Neutron Scattering}
Neutron powder diffraction (NPD) measurements were performed on the high-resolution powder diffractometer HB-2A \cite{18_calder} of the High Flux Isotope Reactor (HFIR) at ORNL in applied magnetic fields up to 4~T and temperatures down to 1.6~K. For these measurements, Pr$_2$Sn$_2$O$_7$ powder was pressed with a Carver press into pellets which were loaded in a cylindrical Cu can. Diffraction data were collected with a neutron wavelength $\lambda_\mathrm{n}$ of 2.41~\AA~and a collimation of open-21$'$-12$'$. Rietveld refinements were performed using the TOPAS ACADEMIC v6 software package and the magnetic structure symmetry analysis was performed using SARAh \cite{00_wills}.

Polarized diffuse neutron scattering experiments were performed on the diffuse scattering spectrometer D7 \cite{09_stewart} at the Institut Laue-Langevin (ILL). An incident wavelength of 4.8~\AA~was selected by a pyrolytic graphite monochromator. Data were collected in the absence of an external DC field and in non-time-of-flight mode, leading to the extraction of the energy-integrated scattering intensity between $-20 \leq E \leq 3.5$~meV. Pr$_2$Sn$_2$O$_7$ powder was loaded in a double-wall cylindrical Cu can and then placed in the dilution fridge insert of a cryostat with a base temperature of 0.05~K. Data were collected at 0.05, 1.0, and 50~K. Data normalization by a vanadium standard ensured that differences in detector efficiency and solid angle coverage were taken into account. Scattering contributions from an empty and a cadmium-filled sample holder were added together and weighted by the sample transmission to estimate the instrument background. Corrections for polarization efficiency of the supermirror analyzers were made by using the scattering from amorphous quartz. Equal counting times were spent on measuring the scattering along the $x$, $y$, and $z$ directions. The 6-pt. $xyz$ polarization analysis method \cite{93_scharpf} was used to separate the magnetic, nuclear coherent, and nuclear-spin incoherent scattering channels for each wavevector transfer magnitude $Q$. The non-spin-flip and spin-flip scattering along each of the three directions were measured with a time ratio of 1:4. It has been shown that the 6-pt. $xyz$ polarization analysis method becomes inaccurate at small $Q$, where a spurious transfer of intensity from the magnetic to the nuclear-spin incoherent channel occurs due to scattering away from the horizontal plane \cite{ehlers_d7}. We performed an approximate correction for this effect by fitting the nuclear spin-incoherent channel at each temperature to the form $c+b/Q^2$, where $b$ and $c$ are fit parameters, and then obtaining the corrected magnetic scattering by adding the $b/Q^2$ term to the uncorrected magnetic scattering.

Low energy inelastic neutron scattering (INS) measurements were collected using the Disk Chopper Time-of-Flight Spectrometer (DCS) \cite{03_copley} at the National Institute of Standards and Technology (NIST) Center for Neutron Research (NCNR) in two separate experiments. Pr$_2$Sn$_2$O$_7$ powder was loaded in a 1.25-cm diameter cylindrical Cu can and then placed in a dilution fridge insert of a cryomagnet achieving a base temperature of 0.2~K in experiment 1 and 0.02~K in experiment 2. Data were collected at 0.2 and 0.5~K in Exp.~1 and at 0.02, 0.5, and 4.2~K in Exp.~2. An incident wavelength of $9$~\AA~and low-resolution mode were chosen, corresponding to a flux of $\sim$ 2 $\times$ 10$^{5}$  neutrons/cm$^2$-s with an elastic line resolution (full-width half-maximum) of approximately 0.018~meV, and an accessible $Q$ range of $[0.1,1.3]$~\AA$^{-1}$ in the elastic channel. The DCS data were corrected for neutron absorption by the sample.

\section{Results and Discussion}

\subsection{Bulk magnetic measurements}
\begin{figure*}
\centering
\scalebox{0.7}{\includegraphics{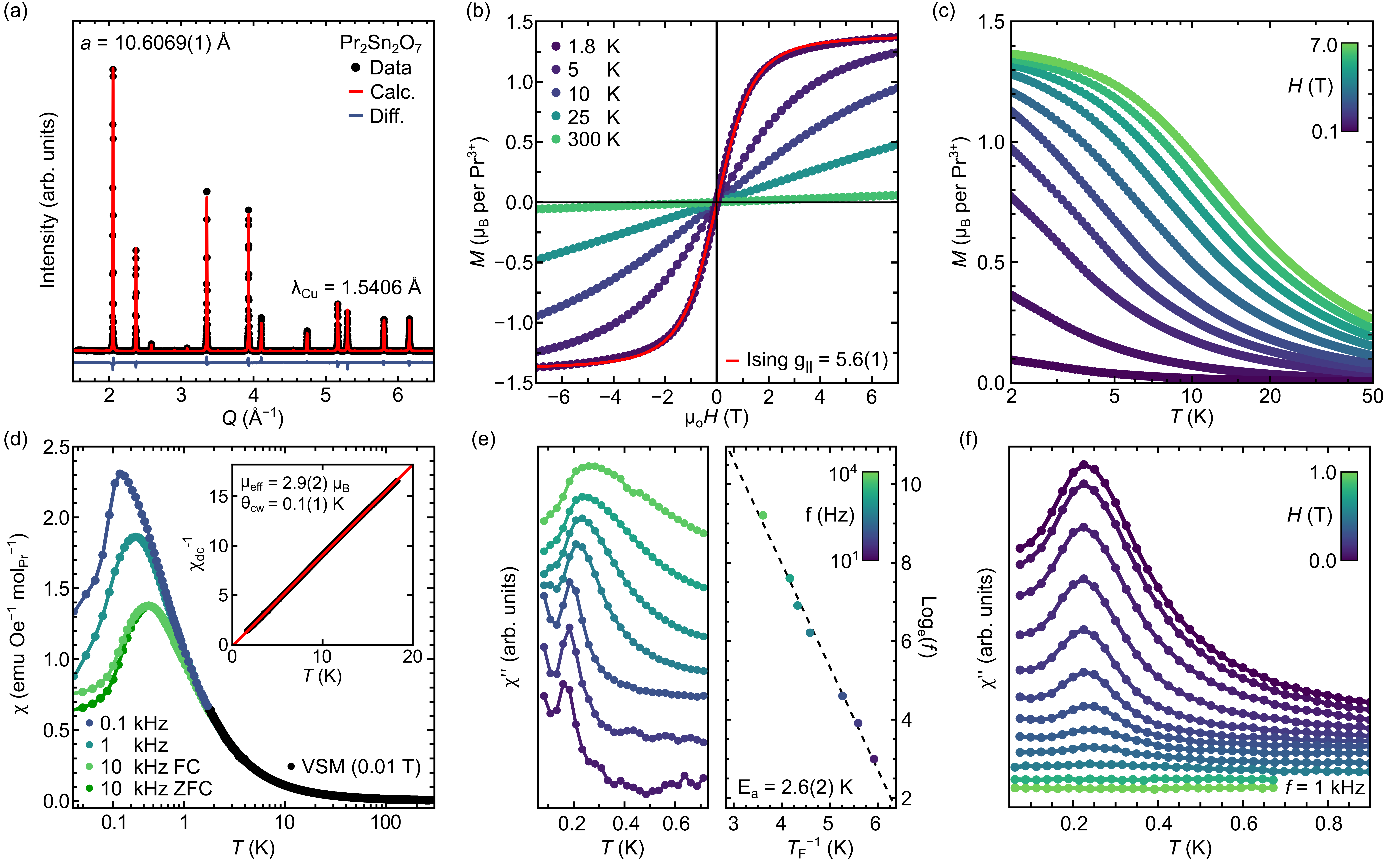}}
\caption{\label{Fig1} (color online) (a) Measured, calculated, and difference room temperature XRD profiles of polycrystalline Pr$_2$Sn$_2$O$_7$. (b) Field-dependence of the isothermal magnetization data with the best fit to the powder-averaged $J{\rm{_{eff}}}=\frac{1}{2}$ Ising model ($g_{\perp}=0$) superimposed on the data. (c) DC (VSM) magnetization measurements as a function of magnetic field. The derivative $dM/dH$ was used to extract the characteristic field for the onset of the field-induced spin-ice order. (d) Temperature dependence of the DC susceptibility and the AC susceptibility at selected frequencies. As a test, the 10~kHz data shows both the FC/ZFC measurements under a 0.05~T field, illuminating a weak history dependence. The inset shows the Curie-Weiss analysis (fit from 3~K to 20~K) of the DC susceptibility data, which suggests the presence of weak net ferromagnetic exchange.  (e) Temperature dependence of the imaginary part of the AC susceptibility $\chi^{\prime \prime}$ for various frequencies in the absence of an external DC field with the corresponding Arrhenius fit of the freezing transition, which yields an activation energy $E_a =$~2.6~K. (f) DC field dependence of the freezing transition in $\chi^{\prime \prime}$ using an intermediate driving frequency $f =$~1~kHz. }
\end{figure*}

The magnetic properties of pyrochlores with non-Kramers crystal field doublets at the rare earth sites have shown remarkable sensitivity to chemical disorder, crystalline defects, and synthetic processes \cite{17_savary, 18_benton, 22_tang,14_koohpayeh,13_kimura}. The potential for mixed valence defects is particularly prevalent in Pr-containing compounds, as non-magnetic Pr$^{4+}$ can be disruptive to potentially fragile magnetic ground states. Previous work on polycrystalline Pr$_2$Sn$_2$O$_7$ highlights the importance of sample dependence, as Ref.~\cite{08_zhou} identifies quasi-elastic scattering that is typically associated with the enhanced low-temperature magnetic monopole density expected for a QSI state \cite{21_jaubert}, while Ref.~\cite{04_matsuhira} presents AC susceptibility data that is more consistent with a CSI state that freezes completely. Since these studies focus on different measurements, direct comparisons between the two samples cannot be made. To provide a comprehensive examination of the properties of polycrystalline Pr$_2$Sn$_2$O$_7$ produced by \textit{standard solid-state methods}, our manuscript examines a full suite of magnetic, thermodynamic, and neutron scattering measurements on traditionally-prepared samples. 

Samples produced \textit{via} classical solid-state methods exhibit excellent crystallinity and purity. Powders produced in this manner have a light red-brown hue, which suggests they likely contain a small concentration of Pr$^{4+}$ defects. Informal discussion with authors of Ref.~\cite{08_zhou} corroborate the coloration of polycrystalline samples. Notably, Pr$^{3+}$ oxides and Pr$^{4+}$ oxides are typically green and red in color respectively, with mixed Pr valence systems taking on darker shades in some cases. In fact, it has been shown that the color of floating-zone grown Pr$_2$Zr$_2$O$_7$ crystals change from dark brown to green as Pr$^{4+}$ defects are suppressed during growth \cite{14_koohpayeh}. Figure~\ref{Fig1}(a) presents the x-ray diffraction pattern for a representative Pr$_2$Sn$_2$O$_7$ powder sample, with the Rietveld refinement superimposed on the data. There are no discernible impurity peaks in the pattern and the lattice constant $a=10.6069(1)$\,\AA~agrees well with previous work \cite{83_subramanian, 97_kennedy, 02_matsuhira, 08_zhou}. Occupancies for Pr and Sn were refined, but did not deviate substantially from unity and were subsequently fixed. 

Figure~\ref{Fig1}(b) depicts the magnetization $M$ measured as a function of applied field $\mu_0 H$ at selected temperatures. The 1.8 K magnetization was fit to the appropriate expression for $J{\rm{_{eff}}}=\frac{1}{2}$ Ising pyrochlores \cite{00_bramwell}, which is overplotted on the data in Fig.~\ref{Fig1}(b) to extract a $g_{zz}$ ($g_\parallel$) value of 5.6(1). Figure~\ref{Fig1}(c) depicts the DC magnetization for Pr$_2$Sn$_2$O$_7$ as a function of temperature and magnetic field. While the low-field data ($\mu_0 H <$~2.5~T) exhibits Curie-Weiss temperature dependence down to 1.8 K, higher-field measurements feature an inflection point in the data that is indicative of a magnetic transition or crossover to a ground state with a net moment. To provide a complete picture of the temperature-dependent magnetization, Fig.~\ref{Fig1}(d) overlays the low-field (0.01~T) DC susceptibility data (1.8-300~K) with the real component of the AC susceptibility (60~mK-4~K). The inset shows the linear behavior of the inverse DC susceptibility data in the temperature range 3-20 K, which was fit to a Curie-Weiss law to extract an effective moment 2.9(2)~$\mu_{\text{B}}$/Pr$^{3+}$ and the ferromagnetic Curie-Weiss temperature 0.1(1)~K. Both values are in good agreement with other studies \cite{02_matsuhira, 08_zhou}. We also show a set of zero-field-cooled (ZFC) and field-cooled (FC) measurements for a high-frequency (10~kHz) AC susceptibility curve under a small (0.05~T) field. The onset of a slight splitting in the FC/ZFC curves is noted, coinciding with the primary peak in the susceptibility. This could suggest a minor contribution from a potentially glassy phase, such that the field history should be taken into consideration. Unless explicitly noted, all further thermodynamic measurements in this manuscript are performed in field-cooled conditions.

The absolute units for the AC susceptibility data were obtained by normalizing the lowest frequency data to the DC susceptibility. There are two features of note in the low-temperature data: 1) a clear onset of spin freezing around 0.1-0.3K, and 2) incomplete freezing as $T \rightarrow 0$. The onset of spin freezing was identified in Pr$_2$Sn$_2$O$_7$ previously \cite{04_matsuhira} and is known in other CSI systems \cite{00_matsuhira, 01_matsuhira}. However, samples in Ref.~\cite{04_matsuhira} exhibit full spin freezing over all frequencies, where $\chi^{\prime \prime} \rightarrow 0$ as  $T \rightarrow 0$. Our samples exhibit a partially frozen state, which is more consistent with recent results in other Pr pyrochlores \cite{13_kimura, 16_petit, 16_sibille}. It has been suggested that complete spin freezing, which has also been identified in a sample of Pr$_2$Hf$_2$O$_7$ \cite{16_anand}, is extrinsic in origin. The wide sample dependence in Pr$^{3+}$-based pyrochlores speaks to the defect-sensitivity of the materials, and one possibility is that the spin freezing is nucleated by defects and/or disorder. We suspect that the low-temperature AC susceptibility signature should be suppressed in pristine samples. While a future goal will be to explore the defect energetics within Pr$_2$Sn$_2$O$_7$, our current manuscript is designed to clarify the state of traditionally prepared Pr$_2$Sn$_2$O$_7$ powders.

\begin{figure*}
\centering
\scalebox{0.7}{\includegraphics{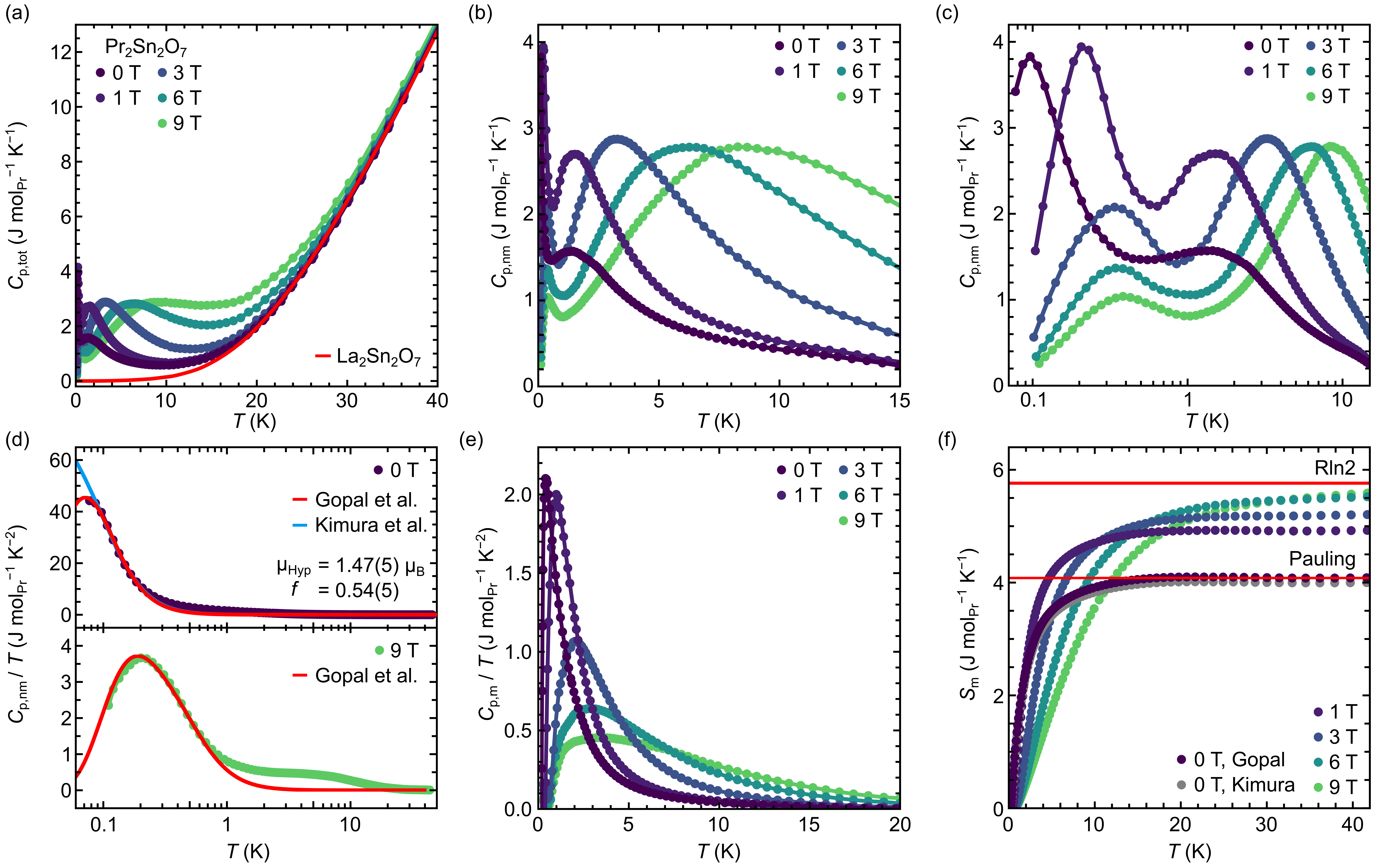}}
\caption{\label{Fig2} (color online) (a) Heat capacity data for polycrystalline Pr$_2$Sn$_2$O$_7$ at selected magnetic fields. The lattice contribution from non-magnetic La$_2$Sn$_2$O$_7$ is indicated by a solid red curve. (b) The lattice-subtracted specific heat as a function of field demonstrates a low-temperature nuclear Schottky anomaly and a higher temperature peak. (c) The same data shown in (b), but now presented using an x-axis log scale to emphasize the two distinct features observed for a given field. (d) Modelling of the nuclear Schottky anomaly using two different models: (1) an analytic form used by Kimura {\it et al.} \cite{13_kimura} in Pr$_2$Zr$_2$O$_7$ and (2) a generalized statistical-mechanical model provided by Gopal {\it et al.} \cite{66_Gopal}. (e) The magnetic component of $C_{p,m}/T$ after removal of the nuclear component. (f) Integration of the magnetic heat capacity provides insight into magnetic ground state. Regardless of the choice of nuclear Schottky model, at zero field we recover the full Pauling entropy, consistent with a spin-ice ground state. At the highest magnetic fields measured here, we nearly recover the full $R\ln(2)$ expected for a well-isolated crystal field ground state doublet.}
\end{figure*}

The frequency dependence of the imaginary part of the AC susceptibility $\chi^{\prime\prime}$ is shown in Fig.~\ref{Fig1}(e), with the freezing transition $T_f$ indicative of slow dynamics defined as the temperature corresponding to the signal's maximum value. The low-frequency value of $T_f =$~0.17 K is significantly suppressed compared to the 1 K energy scale for CSI \cite{00_matsuhira, 01_matsuhira}. The frequency dependence of $T_f$ is well-described by the Arrhenius law $f = f_0 e^{-E_A/T_f}$, with slightly smaller $E_A$ and $f_0$ values of 2.6(2)~K and 110(10)~MHz as compared to previous work \cite{04_matsuhira}. We present the field-dependence of $\chi^{\prime\prime}$ using an intermediate frequency of 1 kHz in Fig.~\ref{Fig1}(f), where we find that the freezing transition temperature is unaffected until the peak is completely suppressed for applied fields $\mu_0 H \ge$~0.75~T.

\subsection{Heat capacity}

Next, we revisit the heat capacity of Pr$_2$Sn$_2$O$_7$ by collecting data down to dilution-refrigerator temperatures in both zero and applied magnetic fields. Our measurements are motivated by two main factors. First, a careful analysis of the nuclear Schottky anomaly contribution can help to constrain the spin correlation time associated with the magnetic ground state of the system. Secondly, an entropy analysis that includes lower temperature data may help to resolve a longstanding puzzle in the literature - previous heat capacity work reports a zero-field entropy that falls substantially short of the Pauling's value for a classical spin ice state \cite{08_zhou}. This result is seemingly at odds with the near-complete spin freezing observed in earlier AC susceptibility measurements \cite{04_matsuhira} and previous work on the zirconate analog Pr$_2$Zr$_2$O$_7$ \cite{13_kimura}. 

Figure~\ref{Fig2}(a) presents heat capacity data ($C_{\text{p,tot}}$) for traditionally-prepared, polycrystalline Pr$_2$Sn$_2$O$_7$ at selected fields with the red solid curve indicative of the lattice contribution for non-magnetic La$_2$Sn$_2$O$_7$. Figure~\ref{Fig2}(b) shows the same data but now with the lattice contribution subtracted, so only the nuclear and magnetic components are remaining ($C_{\text{p,nm}}$). As best observed in Fig.~\ref{Fig2}(c), there are two principal features in the $C_{\text{p,nm}}$ zero-field data: a broad peak centered around 1~K, and a sharper peak close to the base temperature of 60~mK. This two peak structure is also visible in the finite field data, with both peak positions displaying significant field-dependence. The higher-temperature peak signifies the development of spin ice correlations or a transition/crossover from a paramagnetic to an ordered ground state in the low and high-field regimes, respectively. The low-temperature peak has a nuclear Schottky origin, which must be subtracted off from the $C_{\text{p,nm}}$ data to obtain a reliable estimate for the magnetic entropy release of this system.

In general, a Schottky anomaly manifests when there is a system where particles can exist in $m$ energy levels, separated from the ground state by the associated energies $\epsilon_1, \epsilon_2,...\epsilon_m$ with degeneracies $g_1, g_2,...g_m$. A statistical approach using the Boltzmann factor can be used to drive the specific heat of a system of $N$ particles at an effective temperature $T$ \cite{66_Gopal}:
\begin{equation}
C_{n}= \frac{{d}}{{d}T} \left[ \frac{N \sum^m_{r=0} \epsilon_r g_r e^{-\epsilon_r/kT} }{\sum^m_{r=0}  g_r e^{-\epsilon_r/kT}} \right]
\label{sch}
\end{equation}
Note that while the magnetic field is not contained explicitly within Eq.~\ref{sch}, for the special case of a nuclear Schottky anomaly it generates a  splitting of the nuclear energy levels that gives rise to the associated energies $\epsilon_1, \epsilon_2,...\epsilon_m$. There are several types of assumptions that can be made to evaluate the temperature dependence of the nuclear Schottky heat capacity. The most trivial approach is to assume a power series expansion of the specific heat; keeping the first term yields $C_{n} \propto T^{-2}$. This has been applied in many Pr$^{3+}$-based systems and works best when only the tail of the nuclear Schottky anomaly is captured or when the majority of the magnetic contribution is well separated from the nuclear contribution \cite{Tm2_anand2016physical,Tm2_hejtmanek2013phase,Tm2_jin1994kondo,Tm2_pathak2013anomalous,Tm2_sahling1982heat,gopal_Tm2_movshovich1999magnetic}. The second approximation is to assume a multi-level system with evenly spaced energy levels.\cite{gopal_adhikari2019magnetic,gopal_spectro_chirico1979low,gopal_Tm2_movshovich1999magnetic,GopalModel_mitarov1975schottky,GopalModel_vollmer2003low,gopal_matsuhira2011metal,66_Gopal} Here, $\epsilon_i = i \times \epsilon_0$, where $\epsilon_0$ is the characteristic energy splitting. This approach allows Eq.~\ref{sch} to be evaluated numerically and is relatively flexible. A common simplification of this form is the application of a two-level system ($m=1$, $\epsilon_i \in 0, \epsilon_1$). The most general approach would be to use the exact $\epsilon_i$ values derived either from spectroscopic or theoretical approaches, though there is often difficulty generalizing these results, particularly if magnetic fields are involved \cite{gopal_spectro_chirico1979low,spectro_bauer2016schottky}.

Another approach of note is the method utilized by Kimura {\it et al.} in their analysis of Pr$_2$Zr$_2$O$_7$ \cite{13_kimura}, with direct applicability to our data on Pr$_2$Sn$_2$O$_7$. They obtained the following expression for the Pr nuclear spin contribution to the specific heat arising from hyperfine dipole field splitting of the nuclear spin multiplet due to frozen Pr electronic moments \cite{13_kimura}: 
\begin{equation}
C_{n} = \frac{N k_\mathrm{B}\alpha^2}{4 I^2} \left[\frac{1}{\sinh^2(\alpha/2I)} - \frac{(2I+1)^2}{\sinh^2[(2I+1)\alpha/2I]}\right]
\end{equation}
where $\alpha = {A_{\mathrm{ hf}} \mu_{\mathrm {hyp}}^{\mathrm{ Pr}} I}/{g_{\mathrm{J}} k_{\mathrm{B}} T}$, $N$ is Avogadro's number, $k_{\mathrm{B}}$ is the Boltzmann constant, $I =5/2$ and $A_\mathrm {hf}$ are the nuclear spin and hyperfine coupling constant for $^{141}$Pr, and $g_{\mathrm{J}} =4/5$ and $\mu_{\mathrm {Hyp}}^{\mathrm {Pr}}$ are respectively the Lande g-factor and static magnetic dipole moment for Pr$^{3+}$. One advantage of this methodology is the comparison of the absolute magnitude of the observed Schottky anomaly with the theoretical prediction. In Pr$_2$Zr$_2$O$_7$, calculations based on the full ground state Pr$^{3+}$ doublet predict a large Schottky peak of $\sim$7~J/mol~K, which is substantially larger than the values observed here, or in the zirconate analog \cite{13_kimura} (2--3~J/mol~K). Kimura {\it et al.} interpreted this result as signifying that a significant fraction of the Pr sites are not static on the nuclear hyperfine interaction time scale $\tau >$~1 ns. \cite{13_kimura, scheie_thesis}.

For our analysis, we did the following: 1) fit the expression derived by Kimura {\it et al.} to the zero-field $C_{\text{p,nm}}$ data for direct comparison, 2) fit the generalized form given by Gopal {\it et al.} in Eq.~\ref{sch} to both the zero-field and finite field data under the assumption that the energy levels split evenly (i.e. $\epsilon_i = i \times \epsilon_0$). The limiting cases of 0~T and 9~T are clearest. The low-field limit is expected to approach that of a two-level fit and the high-fields at 9~T should be sufficient to fully split adjacent levels. The energy splitting was determined empirically using a least-squares fit to the data. We stress that these approaches are still approximations. However, we ultimately show that both approaches yield consistent results, despite their substantial differences. 

Figure~\ref{Fig2}(d) shows the results of the nuclear Schottky modeling for both the 0~T (top) and 9~T (bottom) data sets. To highlight the Schottky contribution, results are shown in terms of $C_{\text{p,nm}}/T$. For the 0~T data, the results based on Kimura (blue) and Gopal's methods (red) are both shown. There is a small discrepancy between the two models as $T\rightarrow0$~K, but they agree remarkably well for the majority of the temperature-range considered here. Presuming that the lowest-$T$ few data points are dominated by the nuclear Schottky contribution and can be excluded from the Kimura model, both models should yield similar results. The extracted hyperfine moment and the phenomenological scale parameter $f_{Sch.}$ are shown on the plot. The magnitude of the nuclear Schottky peak for our Pr$_2$Sn$_2$O$_7$ samples is approximately 3.8~J/mol~K, which falls short of the predicted 7.06~J/mol~K Schottky peak for a frozen spin system with the full crystal field moment of 2.6~$\mu_{\text{B}}$ \cite{13_princep}. The best fit yields $\mu_{\text hyp}^{\text Pr} =$~1.47(5)~$\mu_{\text{B}}$ and $f =$~0.54(5). The values of $f_{\text{Sch.}} <$~1 and $\mu_{\text hyp}^{\text Pr} <$~2.6~$\mu_{\text{B}}$ may arise from sample in-homogeneity and/or low-lying electronic spin fluctuations, which are likely both contributing factors here. These results are reminiscent of previous findings for the sister compound Pr$_2$Zr$_2$O$_7$ \cite{13_kimura, 22_tang}.

Subtracting the nuclear contributions from the various datasets collected at fixed fields yields the magnetic heat capacity $C_{\text{m}}$, which is shown in Fig.~\ref{Fig2}(e) as $C_{\text{p,m}}/T$. The subtracted data contain smooth, broad magnetic peaks. Integrating the magnetic heat capacity yields the magnetic entropy $S_{\text{m}}$, shown in Fig.~\ref{Fig2}(f). Remarkably, we find that the zero-field data recovers the Pauling spin-ice entropy and does not depend substantially on the chosen nuclear Schottky model. The application of magnetic fields up to 9~T increases the recovered entropy, and appear to approach the expected $R\ln2$ for a well-isolated crystal field ground state doublet. These results are similar to the findings in defect-rich \cite{13_kimura} and pristine \cite{22_tang} Pr$_2$Zr$_2$O$_7$ single crystals, which both also host magnetic ground states with residual Pauling entropy.   

\subsection{High-field neutron diffraction}

\begin{figure}
\centering
\scalebox{0.75}{\includegraphics{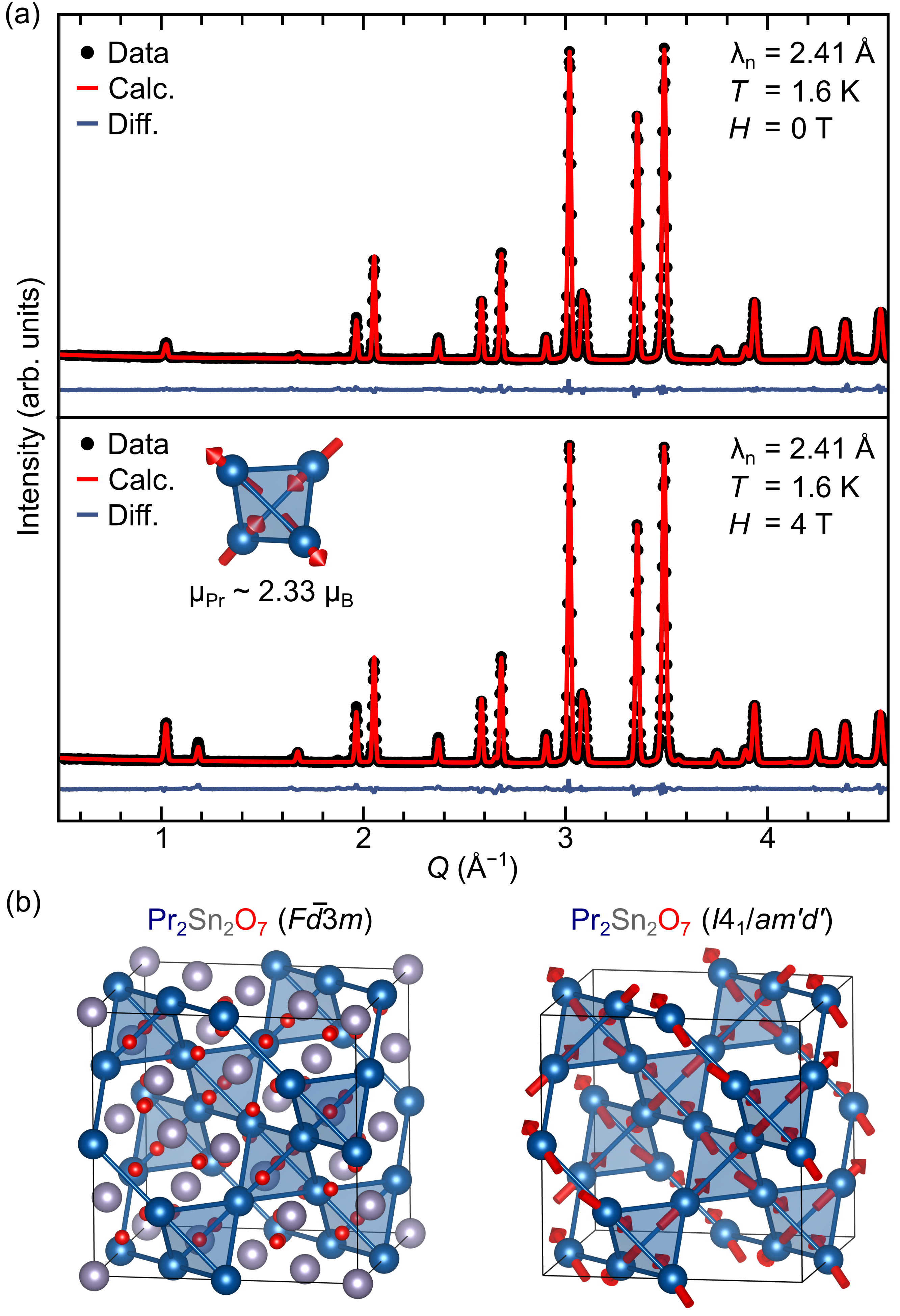}}
\caption{\label{Fig3} (color online) (a) Measured, calculated, and difference neutron powder diffraction profiles ($T =$ 1.6~K, $\lambda_{\mathrm{n}} = 2.41$~\AA) of polycrystalline Pr$_2$Sn$_2$O$_7$ in both the absence and presence of an external DC magnetic field. Contributions from the Bragg reflections of the sample can were fit using a Pawley refinement. Magnetic intensity appears in the 4~T data at $\vec{k} =\vec{0}$ Bragg positions. (b) The parent crystal structure of Pr$_2$Sn$_2$O$_7$ (left), and the field-induced ordered spin ice magnetic structure (right) showing only the Pr atoms for clarity.}
\end{figure}

To characterize the high-field magnetic ground state inferred from our bulk characterization measurements, we collected high-resolution diffraction patterns at temperatures $T \ge$~1.6~K in selected applied fields using HB-2A. The 0 T and 4 T data collected at 1.6~K is shown in Fig.~\ref{Fig3}(a). While the 0 T data refined well using a single phase for the sample corresponding to the expected pyrochlore crystal structure shown in Fig.~\ref{Fig3}(b), the 4 T data included enhanced intensity at $\vec{k} =\vec{0}$ positions and therefore different magnetic models were considered that could account for this additional signal. 

Symmetry analysis using SARAh \cite{00_wills} identified four irreducible representations (IRs) given by $\Gamma_3$, $\Gamma_5$, $\Gamma_7$, and $\Gamma_9$ in Kovalev's notation \cite{93_kovalev}, with detailed information about the possible spin configurations for each IR presented elsewhere \cite{16_hallas}. The magnetic intensity in the 4~T diffraction pattern presented in Fig.~\ref{Fig3}(a) was captured well by a $\Gamma_9$ magnetic structure. The $\Gamma_9$ IR consists of two basis vectors $\psi_1$ and $\psi_2$ that correspond to moments lying parallel and perpendicular to the local $\langle 111 \rangle$ directions respectively, with superpositions of them giving rise to intermediate moment directions. Notably, the model that generates the lowest agreement factor $R_{mag}$ with the 4~T Pr$_2$Sn$_2$O$_7$ data consists of a finite contribution for $\psi_1$ only and an ordered moment of 2.33(4)~$\mu_{\text{B}}$. 

This result agrees relatively well with the value of the crystal field moment associated with the non-Kramers ground state doublet (2.6~$\mu_{\text{B}}$  \cite{13_princep}) and the effective moment extracted from the the low-temperature susceptibility presented in this work (2.9(2)~$\mu_{\text{B}}$). The magnetic structure corresponds to the ordered spin ice phase (No. 141.557, \textit{I}4$_1$/\textit{am}$^\prime$\textit{d}$^\prime$), as shown in Fig.~\ref{Fig3}(b), where all the tetrahedra in the system satisfy the same ice rule. An identical spin configuration is induced by an applied magnetic field $\vec{H} \parallel [001]$ in both ferromagnetic and antiferromagnetic Ising pyrochlores, including Dy$_2$Ti$_2$O$_7$ \cite{02_fennell, 05_fennell}, Ho$_2$Ti$_2$O$_7$ \cite{05_fennell}, Tb$_2$Sn$_2$O$_7$ \cite{05_mirebeau}, and Nd$_2$Ir$_2$O$_7$ \cite{15_ueda}. A zero-field ordered spin ice state has also been identified in pyrochlores with two magnetic sites such as Nd$_2$Mo$_2$O$_7$ \cite{01_yasui} and Sm$_2$Mo$_2$O$_7$ \cite{08_singh}. A field-induced canted version of this phase has been observed in polycrystalline Nd$_2$GaSbO$_7$ \cite{21_gomez} and has been suggested in polycrystalline Nd$_2$InSbO$_7$ \cite{ortiz2022traversing}.

\begin{figure}
\centering
\scalebox{0.75}{\includegraphics{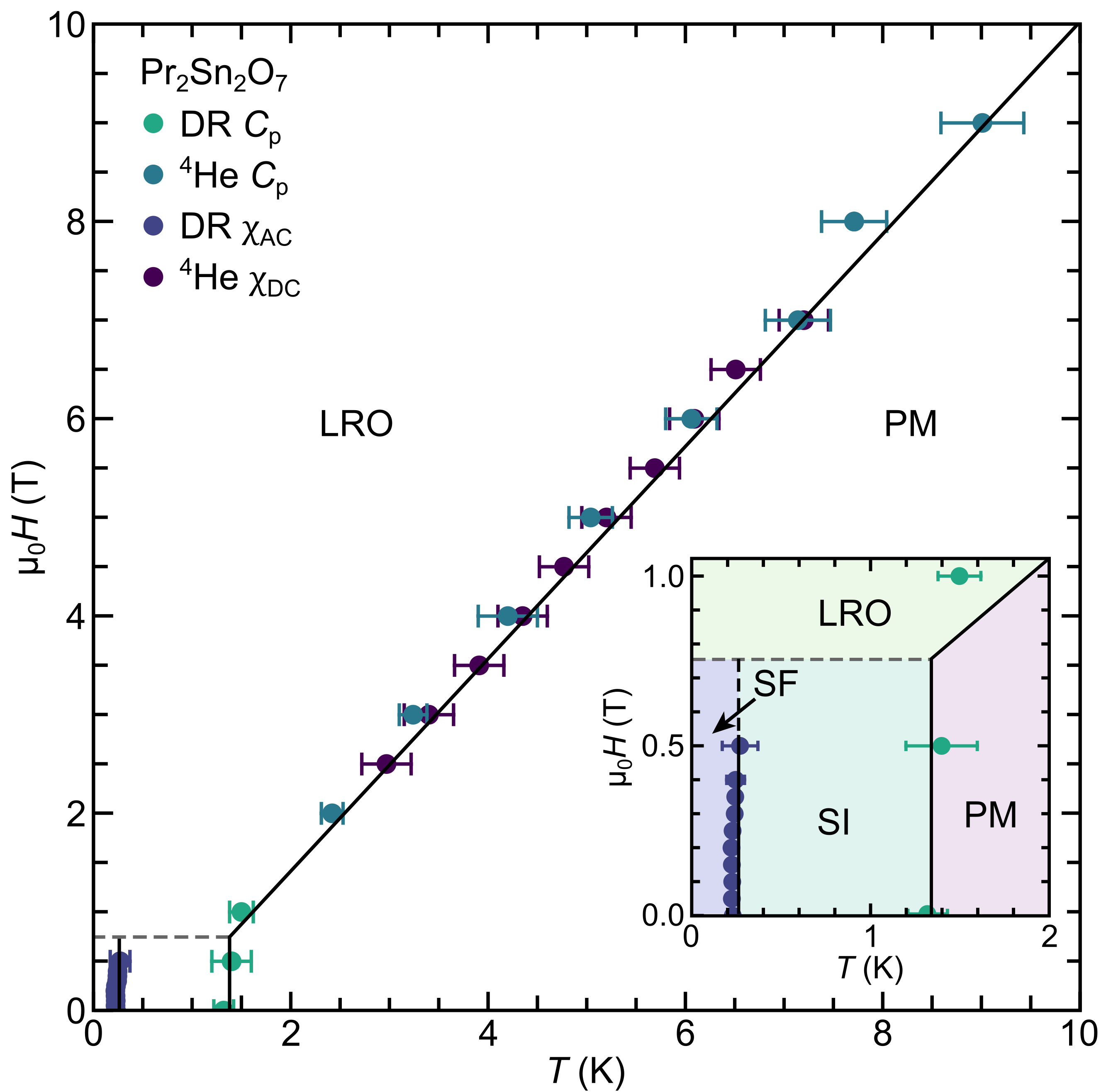}}
\caption{\label{Fig4} (color online) Field-temperature phase diagram of Pr$_2$Sn$_2$O$_7$ as determined by heat capacity (DR $C_{\text{p}}$ and $^4$He $C_{\text{p}}$), AC susceptibility (DR $\chi_{\text{AC}}$) and DC susceptibility ($^4$He $\chi_{\text{DC}}$). Regions include a high-temperature paramagnetic (PM) state, long-range field-induced magnetic order (LRO), spin ice (SI), and partial spin freezing (SF). The inset depicts an enlarged version of the low-temperature, low-field portion of the phase diagram.}
\end{figure}

Figure~\ref{Fig4} summarizes the bulk characterization and HB-2A powder diffraction results for Pr$_2$Sn$_2$O$_7$ in a field-temperature phase diagram, with the different phases labelled as spin ice (SI), field-induced long-range order (LRO), and paramagnetism (PM). The phase boundaries from the heat capacity ($^4$He $C_{\text{p}}$ and DR $C_{\text{p}}$) and DC susceptibility data ($^4$He $\chi_{\text{DC}}$) were determined by the higher-temperature peak position in $C_{\text{p,nm}}$ and the minimum in the first derivative $dM/dH$, respectively. We have also identified the partial spin-freezing (SF) regime, demarcated by the peak position in the imaginary component of the AC susceptibility data (DR $\chi_{\text{AC}}$). Although this partial spin freezing may have an extrinsic origin as described above, it is notably absent in the field-induced ordered state. Tracking the field-dependence of this feature arguably yields the best estimate of the critical field required for the suppression of the ice state, which is given by $\mu_0 H_{\text{c}} =~0.75$~T here.  The fragility of the low-temperature, low-field spin ice state is reminiscent of previous work on the CSI systems Dy$_2$Ti$_2$O$_7$ and Ho$_2$Ti$_2$O$_7$ \cite{05_fennell}, as $\mu_0 H <1$~T applied along the $[001]$ direction was sufficient to generate an ordered spin ice state in those cases. The key features of the phase diagram, including the spin ice state, the low-temperature spin freezing, and the nature of the field-induced order are all hallmarks of a ferromagnetic Ising pyrochlore.

\subsection{Diffuse and inelastic neutron scattering}

\begin{figure}
\centering
\scalebox{0.75}{\includegraphics{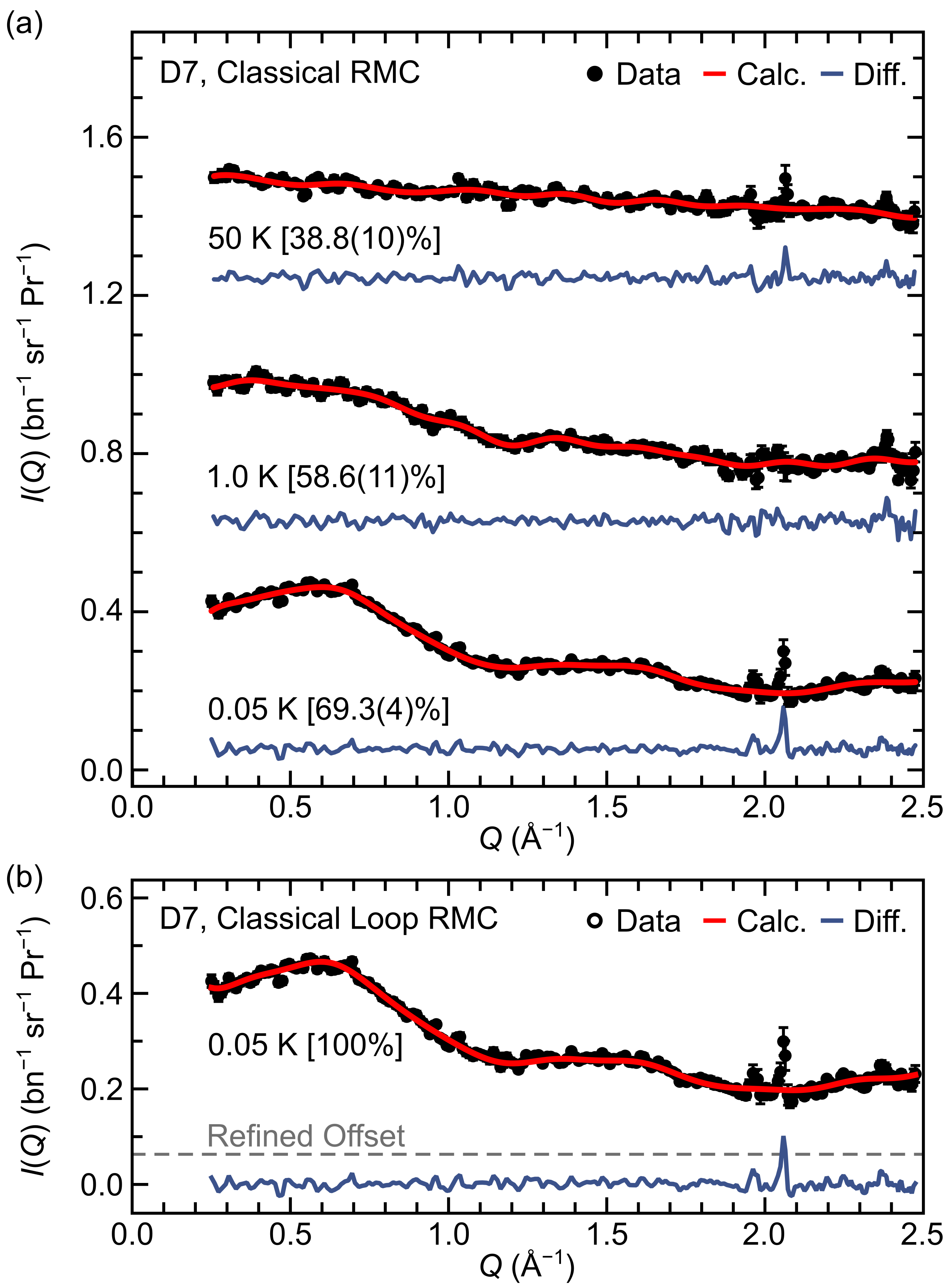}}
\caption{\label{Fig5} (color online) (a) Comparison of the $Q$-dependence of the energy-integrated magnetic scattering for polycrystalline Pr$_2$Sn$_2$O$_7$ at various temperatures with the corresponding reverse Monte Carlo (RMC) fits as described in the text. The data are to scale, but vertically offset for clarity. Percentages denote the portion of tetrahedra that exhibit a two-in-two-out spin configuration. (b) Restricting the available parameter space to lie exclusively in the spin ice manifold for the RMC refinement cannot completely account for the differential magnetic cross section at 0.05~K. Instead, the model requires the addition of a $Q$-independent vertical offset. }
\end{figure}

Finally, we conducted additional neutron scattering measurements to learn more about the properties of the ice state. Magnetic diffuse scattering was identified previously in polycrystalline Pr$_2$Sn$_2$O$_7$ with a $Q$-dependence that was indicative of spin-ice correlations \cite{08_zhou}. While it was initially shown that a CSI model did not provide a complete description of the data, improved agreement between the data and theory was obtained by including the effect of quantum fluctuations in the latter \cite{10_onoda, 11_onoda}. We provide more insight into the nature of this magnetic diffuse scattering by presenting polarized neutron diffraction data from the D7 diffractometer that was analyzed with the reverse Monte Carlo (RMC) method. The results are summarized in Fig.~\ref{Fig5}. The RMC approach  refines magnetic-moment $\vec{S}_i$ configurations directly to experimental data, without making any assumptions about the magnetic Hamiltonian \cite{12_paddison, 13_paddison}. Refinements were performed with the software Spinvert \cite{13_paddison} using $5 \times 5 \times 5$ supercells of the crystallographic unit cell, and initialized with moment orientations randomly assigned as either parallel or antiparallel to the local $\langle 111 \rangle$ axes. At every  iteration of a classical RMC refinement, a single spin is flipped, the goodness-of-fit is calculated according to Ref.~\cite{13_paddison} including a refined intensity scale factor, and the proposed spin flip is accepted or rejected according to the Metropolis algorithm. Refinements were converged until no further improvement in the fit was obtained. The choice of random initial configurations implies that the refined configurations are the most consistent with the experimental data and the assumption of Ising moments.

We performed two different types of refinements. The first method was classical RMC as described above with no additional constraints. Representative fitted curves are superimposed on the data in Fig.~\ref{Fig5}(a), with the percentage of tetrahedra obeying the ice-rules at each temperature indicated on the panel. While the ice correlations are significantly enhanced at low temperature as compared to expectations for a paramagnetic state (i.e. $6/16= 37.5$\% of tetrahedra in the ice configuration), the 69.3(4)\% fraction of tetrahedra obeying the ice rules at 0.05~K is well below the 100\% CSI value. This deviation may arise from local quantum effects, which can cause fluctuations away from the ice-rules manifold. Since the RMC refinements consider magnetic moments as  classical vectors, this potential origin is not distinguishable from thermally-induced spin flips. Another possible explanation is that chemical disorder (e.g. Pr$^{4+}$ defects) could nucleate tetrahedra that do not obey the ice rule. To check this result, the second refinement method (``classical loop RMC") constrained the spin configurations to remain entirely within the ice-rules manifold and fitted the data by using only ``loop moves" that cycle between different ice-rules-obeying states \cite{Hallas_2013}. 

The only way that the magnetic diffuse scattering data could be captured with a fully ice-rules-obeying state was to refine a constant intensity offset as well as an intensity scale factor. A representative fitted curve is superimposed on the 0.05~K data in Fig.~\ref{Fig5}(b), along with the value of this offset. Since our polarized-neutron data should not include any appreciable  contributions from non-magnetic or background scattering, the requirement to include an intensity offset in our classical loop RMC fits provides additional support for the presence of local excitations out of the ice-rules manifold, which would produce a flat contribution to the magnetic scattering.

Neutron spectroscopy measurements obtained using the DCS instrument enable us to discriminate between the elastic and inelastic scattering features of the spin-ice state. Two experiments were performed with the same sample environment and $\lambda_\mathrm{n} = 9$~\AA; the base temperature was 0.20 K for Exp. 1 and and 0.02 K for Exp. 2. Except where otherwise noted, data were collected after cooling the sample in zero applied field.
To remove background scattering, we subtracted data sets measured at base temperature in high applied magnetic fields (6~T for Exp. 1 and 9~T for Exp. 2). Our magnetic field-temperature phase diagram (Figure~\ref{Fig4}) indicates that these parameter combinations place the sample deep within the field-polarized state, where magnetic scattering in the accessible energy range is localized to elastic magnetic Bragg peaks. Hence, the background-subtracted signal away from the Bragg positions provides a good estimate of the purely magnetic scattering. 

\begin{figure*}
\centering
\scalebox{0.7}{\includegraphics{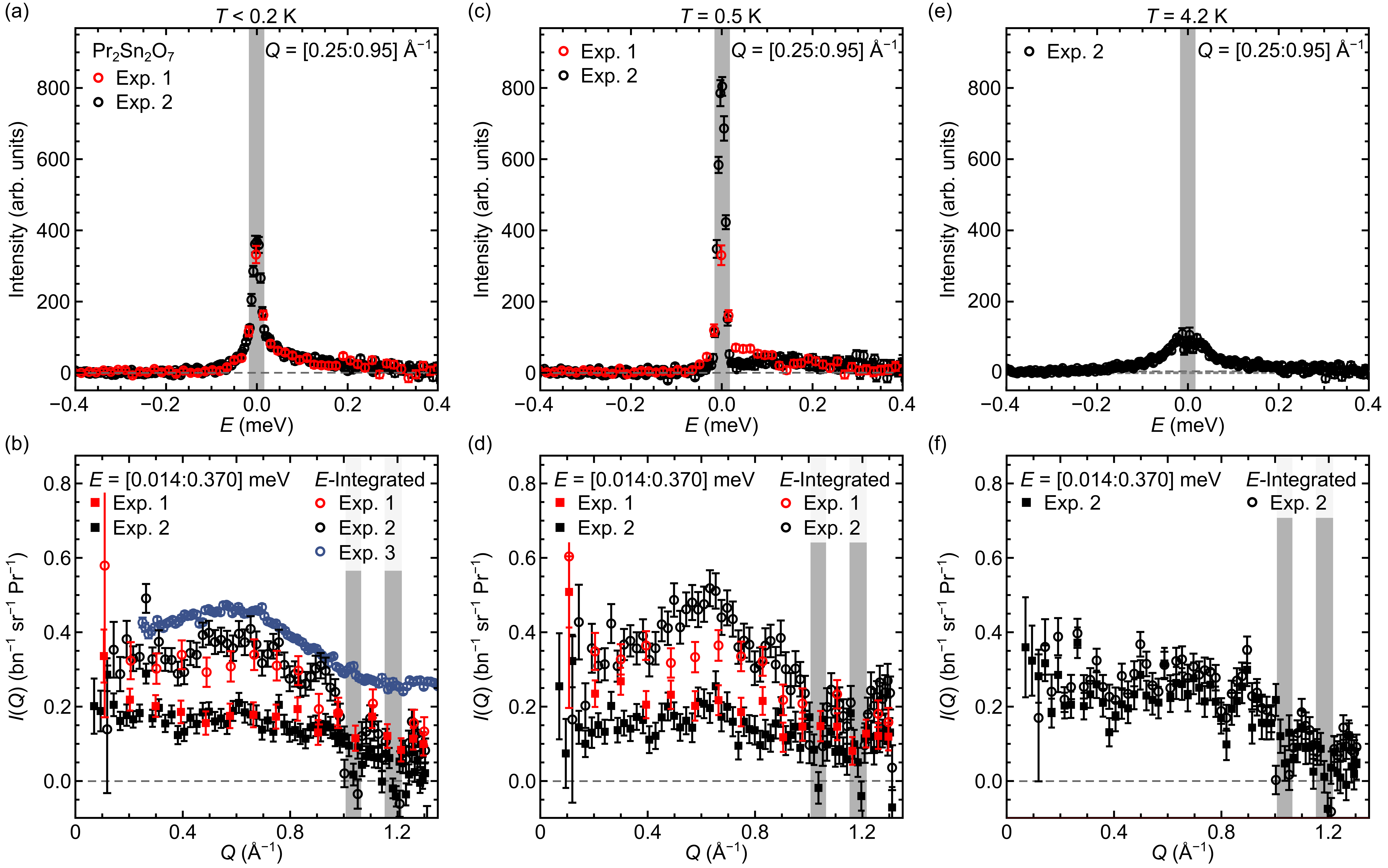}}
\caption{\label{Fig6} (color online) (a) Intensity \emph{vs.} energy transfer $E$ integrated over a $Q$-range of [0.25, 0.95]~\AA$^{-1}$, measured after zero-field cooling at base temperature [$T =0.20$~K for Exp. 1 (red circles), $T =0.02$~K for Exp. 2 (black circles)]. The shaded grey region indicates the elastic portion of the spectrum. (b) Intensity \emph{vs.} momentum transfer $Q$ for energy-integrated data (empty circles), and  inelastic data (filled squares). Data from experiments 1, 2 and 3 are labelled on the plots. For Exp. 1 and Exp. 2, energy integration ranges are $[-0.17,0.37]$~meV and $[0.014,0.37]$~meV for energy-integrated and inelastic data, respectively. For Exp.~3 (D7 data), energy integration extends to $\sim$3.5~meV.
The shaded grey regions denote $Q$-regions affected by over-subtracted  Bragg scattering. (c) Intensity \emph{vs.} $E$ measured in zero field at 0.5~K after zero-field cooling (Exp. 1, red circles), and at 0.5~K in zero field after a 0.02~K measurement in a 9-T field (Exp. 2, black circles), showing a strong dependence on magnetic-field history. (d) Intensity \emph{vs.} $Q$ for energy-integrated data (empty circles) and inelastic data (filled squares). Red symbols show data collected at 0.5~K after zero-field cooling (Exp.~1) and black symbols show data collected at 0.5~K in zero field after a 0.02~K measurement in a 9-T field. (e) Same as (a), except at 4.2~K. (f) Same as (b), except at 4.2~K and with integration range $[-0.37,0.37]$~meV for the energy-integrated data.}
\end{figure*}

Figure~\ref{Fig6}(a) presents the $E$-dependence of the background-subtracted magnetic excitation spectrum at base temperature and in zero magnetic field. These data were integrated over a wide $Q$ range of [0.25, 0.95]~\AA$^{-1}$.  
The quasi-elastic scattering observed previously up to 0.3~meV energy transfers \cite{08_zhou} that has been associated with confined magnetic monopoles \cite{17_sarte} is apparent again here. The DCS data collected in Exp. 1 and Exp. 2 agree within statistical errors, confirming that these results are reproducible.

Figure~\ref{Fig6}(b) presents the $Q$-dependence of the energy-integrated and inelastic components of our base-temperature, zero-field DCS data. We estimated the energy-integrated scattering $I_{\mathrm{tot}}(Q)$ by integrating $I(Q,E)$ over $[-0.17,0.37]$~meV, and the inelastic scattering $I_{\mathrm{inel}}(Q)$ by integrating $I(Q,E)(1+n(E))$ over $[0.014,0.37]$~meV, where $n(E)$ is the Bose factor. The lower bound of the latter integral was chosen because $\sim$93\% of elastic scattering appears at energies below $0.014$~meV. 
These data were converted into absolute intensity units by normalizing to the Bragg intensity of the $(111)$ nuclear peak. As a consistency check, we also plot our equivalently-normalized D7 data as Exp.~3 in Fig.~\ref{Fig6}(b). The $Q$-dependence of the DCS data agrees well with the D7 data, except the overall intensity of the D7 data is slightly higher. This small discrepancy may be due to systematic errors in the normalization procedure, or to the wider $E$-integration range of the D7 data, which extends to $\sim$3.5~meV and so may capture additional high-energy magnetic signals.

We compare the total magnetic moment with the dynamic magnetic moment using the total-moment sum rule \cite{18_sarte, 21_sarte},
\begin{equation}
\mu^{2}_{\rm{{eff}}} = 3 \left(\frac{2}{\gamma r_0}\right)^2 \frac{\int \frac{Q^{2}}{|f(Q)|^2} I(Q) dQ}{\int Q^{2}dQ} ,\label{totmom} 
\end{equation}
where $\left(\frac{2}{\gamma r_0}\right)^2$ is 13.77 sr b$^{-1}$, $f(Q)$ is the Pr$^{3+}$ isotropic magnetic form factor, and $\mu\rm{_{eff}}$ is the magnetic moment. To obtain the total and dynamic moment, respectively $I(Q)=I_{\mathrm{tot}}(Q)$ and $I_{\mathrm{inel}}(Q)$ in Eq.~\ref{totmom}. In principle, the integral in Eq.~\ref{totmom} runs over all $Q$, but in practice, it is necessary to restrict it to our usable $Q$ range of [0.25, 0.95]~\AA$^{-1}$. This approximation is justified here due to broadness of the magnetic scattering features. 

An important quantity is the fraction of the magnetic spectral weight that is dynamic, $(\mu_{\mathrm{eff}}^{\mathrm{inel}}/\mu_{\mathrm{eff}}^{\mathrm{}})^{2}$. Neglecting the effect of finite energy resolution, this fraction would be unity for a fully-fluctuating state, and $1/(J_{\mathrm{eff}}+1)$ for a fully-ordered or frozen magnetic state with quantum number $J_{\mathrm{eff}}$. At 0.02~K, we obtain total and dynamic moments of 2.61(3) $\mu_{\text{B}}$ and 1.80(3) $\mu_{\text{B}}$, respectively, indicating that 48(2)\% of the magnetic spectral weight is dynamic. This value is less than the 67\% expected for a fully-frozen state with $J_{\rm eff} =1/2$ \cite{16_ross, 19_plumb}. Due to the finite energy resolution of our neutron data, we cannot determine how the remaining $\sim$50\% of the spectral weight is distributed between purely elastic scattering ($E=0$) and low-energy quasi-elastic scattering ($0<E<0.014$~meV); hence, our neutron data are consistent with either a fully-frozen or partially-frozen ground state. However, our analysis reveals that a substantial fraction $\sim$50\% of the spectral weight at 0.02~K involves spins with long correlation times $\tau>0.3$~ns.

\begin{table}
\begin{tabular}{cc|ccc}
\hline 
$T$ (K) & Field history & $\mu_{\mathrm{eff}}$ ($\mu_{\mathrm{B}}$) & $\mu_{\mathrm{eff}}^{\mathrm{inel}}$ ($\mu_{\mathrm{B}}$) & $(\mu_{\mathrm{eff}}^{\mathrm{inel}}/\mu_{\mathrm{eff}}^{\mathrm{}})^{2}$\tabularnewline
\hline 
\hline 
$0.02$ & ZF & $2.61(3)$ & $1.80(3)$ & $0.48(2)$\tabularnewline
$0.2$ & ZF & $\mathit{2.52(7)}$ & $\mathit{1.88(7)}$ & $\mathit{0.55(5)}$\tabularnewline
$0.5$  & ZF & $\mathit{2.48(8)}$ & $\mathit{1.97(8)}$ & $\mathit{0.63(7)}$\tabularnewline
$4.2$ & ZF & $2.37(3)$ & $2.14(4)$ & $0.82(4)$\tabularnewline
$0.5$ & $9$ T at $0.02$ K & $2.82(4)$ & $1.79(5)$ & $0.40(2)$\tabularnewline
\hline 
\end{tabular}

\caption{\label{moments}Values of the total magnetic moment $\mu_{\mathrm{eff}}$, dynamic magnetic moment $\mu_{\mathrm{eff}}^{\mathrm{inel}}$, and ratio $(\mu_{\mathrm{eff}}^{\mathrm{inel}}/\mu_{\mathrm{eff}}^{\mathrm{}})^{2}$ obtained from inelastic neutron scattering data at different temperatures and for different magnetic-field histories. Values shown in \emph{italics} were obtained from Exp. 1 and other values from Exp. 2 (see Fig.~\ref{Fig6}).}
\end{table}

We now consider the dependence of the magnetic spectral weight on temperature and magnetic-field history. Our results are summarized in Table~\ref{moments}. Neutron spectroscopy data collected at 0.5~K and 4.2~K are shown in Fig.~\ref{Fig6}(c--f). The total moment obtained varies between 2.4 and 2.8~$\mu_{\text{B}}$, in good agreement with the values obtained from bulk susceptibility and crystal-field analysis \cite{13_princep}. For the data collected under zero-field-cooling (ZFC) conditions, the dynamic fraction of spectral weight increases from $48(2)$\% at 0.02~K to $82(4)$\% at 4.2~K, consistent with increasing thermally-induced spin flips.  At 0.5~K only, data were collected with two different field histories. In Exp.~1, the protocol was ZFC to 0.2~K, measure, warm to 0.5~K, and measure the data shown in red in Fig.~\ref{Fig6}(c,d). In Exp.~2, the protocol was ZFC to 0.02 K, ramp field to 9~T, measure, ramp field to zero, warm to 0.5 K, and measure the data shown in black in Fig.~\ref{Fig6}(c,d). Interestingly, at 0.5~K, the field history of the sample has a significant effect on its magnetic excitation spectrum. For 0.5-K data from Exp.~1, the dynamic fraction of spectral weight is 63(7)\%, but this fraction dramatically reduces to 40(2)\% for data collected in Exp.~2. The low-energy ($E\lesssim0.05$~meV) quasielastic scattering apparent in Exp.~1 appears to collapse to the elastic line in Exp.~2, while higher-energy inelastic scattering remains in Exp.~2. The strong dependence of the magnetic scattering on field history is consistent with the FC/ZFC splitting observed in our bulk magnetic susceptibility data [Fig.~\ref{Fig1}(d)] and is suggestive of glassy dynamics associated with partial spin freezing.

\section{Discussion}

Our combined AC susceptibility, heat capacity, polarized neutron scattering, and neutron spectroscopy results paint an interesting picture of the spin ice state in Pr$_2$Sn$_2$O$_7$. Hallmarks of a homogeneous spin system with slow dynamics required for a CSI state are not apparent here, as the low-temperature drop in the real component of the AC susceptibility data and the nuclear Schottky peak amplitude in the heat capacity data are significantly suppressed as compared to expectations for a fully frozen spin system. These findings are corroborated by the reduced fraction of Pr tetrahedra obeying the ice rules at low-temperature as revealed by the RMC analysis of our magnetic diffuse scattering data. Taken together, these results establish significant spin inhomogeneity with two main components. The slow component has spin correlation times longer than the maximum value detectable by our AC susceptibility measurements (i.e. 0.1 s) at low temperatures. Conversely, we can constrain the spin correlation time of the fast spin component due to the different time scales probed by AC susceptibility, nuclear spin contributions to heat capacity (i.e. spin-lattice nuclear relaxation), and neutron spectroscopy. Since the results from the latter technique are consistent with a fully frozen spin system while the data from the first two techniques are not, this indicates that the spin correlation time for the fast component may straddle the detection limits for the nuclear hyperfine interaction time scale and neutron spectroscopy -- yielding a value on the order of 0.1~ns. Our work suggests that Pr$_2$Sn$_2$O$_7$ has a partially frozen magnetic ground state with a sample-dependent frozen spin fraction nucleated by defects and/or disorder.  

It is interesting to compare our results on Pr$_2$Sn$_2$O$_7$ to previous work on the other Pr pyrochlores. Of these, Pr$_2$Zr$_2$O$_7$ has received the most attention, likely due to the earliest availability of large single crystals. The initial samples were affected by structural disorder likely due to Pr off-centering or Pr$^{4+}$ defects \cite{14_koohpayeh}. On the one hand, these Pr$_2$Zr$_2$O$_7$ samples share some key features with our polycrystalline Pr$_2$Sn$_2$O$_7$ samples, including a partial spin freezing signature in AC susceptibility measurements and a nuclear Schottky amplitude much lower than expected for a fully-frozen spin system \cite{13_kimura}. On the other hand, the spin correlation time for the fast spin component in Pr$_2$Zr$_2$O$_7$ was shorter, as the total moment sum rule result obtained using 0.1~K neutron spectroscopy data revealed predominantly inelastic contributions in this case. The slower spin dynamics identified here in Pr$_2$Sn$_2$O$_7$ may be related to an enhanced amount of disorder and/or defects in the polycrystalline samples as compared to the early Pr$_2$Zr$_2$O$_7$ single crystals. Evidence for a dynamical ground state in Pr$_2$Hf$_2$O$_7$ has also been presented \cite{18_sibille} despite complete \cite{16_anand} or partial spin freezing \cite{16_sibille} signatures in AC susceptibility measurements, but unfortunately dilution fridge heat capacity measurements and a total moment sum rule analysis are not yet available to facilitate comparisons with our work. The collective results on the Pr pyrochlores continue to support their quantum spin ice candidacy, but point to sample-dependent defects and disorder as key issues that need to be better understood and quantified before their intrinsic properties can be revealed. 

\section{Conclusions}
We performed a detailed characterization study to investigate the low-temperature magnetic properties of the ferromagnetic Ising pyrochlore Pr$_2$Sn$_2$O$_7$. We first identify key spin-ice signatures in our polycrystalline samples including a ferromagnetic Curie-Weiss temperature, magnetic moments with Ising anisotropy, and the characteristic magnetic diffuse scattering pattern. Next, we establish a magnetic field-temperature phase diagram for Pr$_2$Sn$_2$O$_7$ that includes all the characteristic features for a ferromagnetic Ising pyrochlore and shows that the spin ice state is sensitive to external perturbations with a critical field of only 0.75~T required to suppress it. Finally, we find important deviations from CSI physics that show the magnetic ground state of Pr$_2$Sn$_2$O$_7$ is partially frozen. More specifically, a finite AC susceptibility signal persists down to the base temperature of 60~mK, and both the amplitude of the nuclear Schottky anomaly and the fraction of Pr tetrahedra obeying the ice-rules are significantly reduced compared to expectations for CSI. It is likely that defect formation (e.g. Pr$^{4+}$) belies the strong sample dependence in Pr$_2$Sn$_2$O$_7$. Regardless, our work suggests that Pr$_2$Sn$_2$O$_7$ remains a strong candidate to host a QSI ground state and calls for renewed efforts to characterize and control defects and disorder in Pr pyrochlores so the intimate connection between their chemistry and magnetic ground states can be definitively established.

\section{acknowledgments}
\footnote{Notice: This manuscript has been authored by UT-Battelle, LLC under Contract No. DE-AC05-00OR22725 with the U.S. Department of Energy. The United States Government retains and the publisher, by accepting the article for publication, acknowledges that the United States Government retains a non-exclusive, paid-up, irrevocable, world-wide license to publish or reproduce the published form of this manuscript, or allow others to do so, for United States Government purposes. The Department of Energy will provide public access to these results of federally sponsored research in accordance with the DOE Public Access Plan (http://energy.gov/downloads/doe-public-access-plan).}

In memoriam, M.W.S. Brown.

B.R.O., A.F.M., and J.A.M.P. gratefully acknowledge support from the U.S. Department of Energy (DOE), Office of Science, Basic Energy Sciences, Materials Sciences and Engineering Division. 
P.M.S. acknowledges additional financial support from the CCSF, RSC, ERC, and the University of Edinburgh through the GRS and PCDS. C.R.W. acknowledges financial support from the CRC (Tier II) program, CIFAR, CFI and NSERC. 
A portion of this research used resources at the High Flux Isotope Reactor, which is a DOE Office of Science User Facility operated by Oak Ridge National Laboratory. 
S.D.W., G.P., M.J.K., S.J.G.A., and P.M.S. acknowledge support from the U.S. Department of Energy (DOE), Office of Basic Energy Sciences, Division of Materials Sciences and Engineering under Grant No. DE-SC0017752.
Neutron data collection (https://doi.org/10.5291/ILL-DATA.5-42-434) on the Diffuse Scattering Spectrometer D7 at the ILL took place with financial support from proposal 5-42-434 awarded to P.M.S. and C.R.W. 
A portion of this work used facilities supported via the UC Santa Barbara NSF Quantum Foundry funded via the Q-AMASE-i program under award DMR-1906325.
\bibliography{PSO_field_paper}

\end{document}